\documentclass[10pt]{article}

\usepackage{amsmath}
\usepackage{amssymb}
\usepackage{placeins}

\usepackage{graphicx}

\usepackage{cite}

\usepackage{color} 
\usepackage{url}
\usepackage{soul}

\usepackage[labelfont=bf,labelsep=period,justification=raggedright]{caption}

\makeatletter
\renewcommand{\@biblabel}[1]{\quad#1.}
\makeatother

\date{}

\begin{document}

\begin{flushleft}
{\Large
\textbf{Early Prediction of Movie Box Office Success based on Wikipedia Activity Big Data}
}
\\
M\'arton Mesty\'an$^{1}$, 
Taha Yasseri$^{1,2,3,\ast}$, 
J\'anos Kert\'esz$^{1,3,4}$
\\
\bf{1} Institute of Physics, Budapest University of Technology and Economics, Budapest, Hungary
\\
\bf{2} Oxford Internet Institute, University of Oxford, Oxford, UK
\\
\bf{3} Department of Biomedical Engineering and Computational Science, Aalto University, Aalto, Finland
\\
\bf{4} Center for Network Science, Central European University, Budapest, Hungary
\\
$\ast$ E-mail: yasseri@oii.ox.ac.uk
\end{flushleft}

\section*{Abstract}
Use of socially generated ``big data'' to access information about  
collective states of the minds in human societies has become a new paradigm in 
the emerging field of computational social science. A natural 
application of this would be the prediction of the society's reaction to a new product in the sense of
popularity and adoption rate. However, bridging the gap between ``real time monitoring'' and ``early predicting'' remains a big challenge. 
Here we report on an endeavor to build a 
minimalistic predictive model for the financial success of movies based on collective activity data of online users. We show that 
the popularity of a movie can be predicted much before its release by measuring 
and analyzing the activity level of editors and viewers of the corresponding
entry to the movie in Wikipedia, the well-known online encyclopedia. 
  
\section*{Introduction}
Living in the digital world of today, along with all the advantages also has its side 
effects and byproducts. Our daily life nowadays leaves a digital trace of all our activities
in the recently developed Information and Communications Technology based environments. Our social
communications through different digital channels, financial activities within e-commerce, physical locations 
registered by cell phone providers etc., are traced and recorded. In addition to such 
passive collection of data about online activity, we also actively share information about our feelings, emotional moods, opinions and views 
through the so called Web~2.0. or user generated content within social media. 
In addition to providing us with novel answers to classic questions about individual and social aspects 
of human life from scientific point of view, precise analysis of this 
huge amount of data can have practical applications to predict, monitor, and cope
with many different type of events, from simple matters of daily life to massive crises in 
the global scale. 
For example, Sakaki et al. have developed an alerting 
system based on Tweets (posts in the Twitter microblogging service),
being able to detect
earthquakes almost in real time \cite{Sakaki2010}.
They elaborate their detection system further to detect rainbows 
in the sky, and traffic jams in cities \cite{Okazaki2011}. The practical
point of their work is that the alerting system could perform so promptly that
the alert message could arrive faster than the earthquake waves to certain regions.
Bollen et al. have analyzed moods of Tweets 
and based on their investigations they
could predict daily up and down changes in  
Dow Jones Industrial Average values with an 
accuracy of 87.6\% \cite{Bollen2011}. 
Saavedra et al. investigated the relationship between the
content of traders' messages and market dynamics. They show that there is a positive 
correlation between the usage of ``bundles'' of positive and negative words
with agents' overall financial performance \cite{Saavedra2011b}.
Another example is using Twitter to predict electoral outcomes \cite{Tumasjan10}, 
however with its biases and limitations \cite{GayoAvello11,GayoAvello2012}.
Interesting studies have appeared treating the use of social media indicators 
to predict the scientific impact of research articles, e.g., short-term web usage 
(number of downloads from the pre-print sharing web site ``arXiv'') \cite{Brody2006}
and Twitter mentions \cite{Eysenbach2011}. In a recent work, 
it is shown that Twitter mentions and arXiv downloads  
follow two distinct temporal patterns of activity, however, the volume of Twitter mentions is
statistically correlated with arXiv downloads and early citations \cite{Shuai2012}.
Preis et al. found a correlation between weekly transaction volumes of ``S\&P 500 companies''
and weekly Google search volumes of corresponding company names \cite{Preis10}. 
By analyzing search queries for information about preceding and following years,
a ``striking'' correlation between a country's GDP and the predisposition of 
its inhabitants to look forward is observed \cite{Preis12}.
Based on Google search logs, Ginsberg et al. estimated the spread of 
influenza in the United States \cite{Ginsberg09}. There are other examples of using social media streams to
make predictions on news popularity in terms of the number of user-generated comments \cite{Tsagkias2009,Tsagkias2010} or the number of news visitors \cite{Castillo2013}. 
For a comprehensive literature review see \cite{tsagkias2012-thesis}.

Statistical analysis of motion picture markets has led to intriguing results, such 
as observing the evidence for a Pareto law for movie income \cite{Sinha2004,Sinha2005} along with
a log-normal distribution of the gross income per theater and a bimodal 
distribution of the number of theaters in which a movie is shown \cite{Pan2010}.
By analyzing historical data about 70 years of the American movie market, Sreenivasan has argued that the movies with higher level of novelty (assigned based on 
keywords from the Internet Movie Database) produce larger revenue \cite{Sreenivasan2013}.
Despite much effort with different approaches, predicting the financial success 
of a movie remains a challenging open problem. 
For example, Sharda and Delen have trained a neural network 
to process pre-release data, such as quality and popularity variables, 
and classify movies into nine categories according to their 
anticipated income, from ``flop'' to ``blockbuster''. 
For test samples, the neural network classifies only 36.9\% of the movies 
correctly, while 75.2\% of the movies are at 
most one category away from correct \cite{Sharda06}. Joshi et al. have built a 
multivariate linear regression model that joined meta-data 
with text features from pre-release critiques to predict the revenue with a coefficient of 
determination $R^2 = 0.671$ \cite{Joshi10}. Since predictions based on classic quality factors fail 
to reach a level of accuracy high enough for practical application, usage of user-generated data to predict the success of a movie
becomes a very tempting approach.
Ishii et al. present a mathematical framework for the spread of popularity in society \cite{Ishii12}. 
Their model, which takes the advertisement budget
as an input parameter and generates a dynamic popularity variable, 
is validated against the number of blog posts on the particular movies in the Japanese Blogosphere. In other words they 
consider the activity level of bloggers as a representative parameter for social popularity. In an earlier work \cite{Hidalgo2006} a quantitative model based on 
``word of mouth'' spreading mechanism was introduced in order to assess the quality of movies based on the ``aggregated consumption data''.
However, by analyzing the sentiment of blog stories on movies, 
Mishne and Glances emphasize that the correlation between pre-release sentiment
and sales is not at an adequate level to build up a predictive model \cite{Mishne2006}.
In a very interesting approach Asur and Huberman set up a prediction system for 
the revenue of movies based on the volume of Twitter mentions \cite{Asur10}. They 
achieve an adjusted coefficient of 
determination of 0.97 on the night before the movie release for the first weekend 
revenue of a sample of 24 movies. 
In a later work, however, Wong et al. show that Tweets do not necessarily represent the 
financial success of movies \cite{Wong2012}. They consider a sample of 34 movies 
and compare the Tweets about the movies to evaluations written by users of movie review web sites. 
They argue that predictions based on social media could have high precision but low recall. 
Yun and Gloor showed that the betweenness centrality of a movie in a network representation of its presence on the Web is correlated with its financial success
\cite{Yun2012}. In a rather novel approach, Oghina et al. have made use of Twitter and YouTube activity streams to predict the 
ratings in the Internet Movie Database (IMDb), which is among the most popular online movie databases \cite{Oghina2012}.

Wikipedia, as a predominant example of user-generated media, 
has been intensely studied from different points of view. 
Its size and growth \cite{voss2005,almeida2007,suh2009}, 
topical coverage and notability of entries \cite{holloway2007,halavais2008,taraborelli2010}, 
conflict and editorial wars among users \cite{sumi2011a,sumi2011b,yasseri2012b,yasseri2012d,torok2012}, 
editorial patterns \cite{yasseri2012a} and 
linguistic features \cite{yasseri2012c} are only few examples of 
research topics associated with Wikipedia.
We are aware of two comprehensive reviews \cite{nielsen2011,jullien2012} and a brief hands-on 
guide to some of the most recent Wikipedia research \cite{yasseri2012e}.

Although effects of external events on the activity of Wikipedia 
editors \cite{Keegan11,ratkiewicz2010} and the number of page views \cite{Spoerri2007a,Spoerri2007b} 
have been studied in detail, usage of Wikipedia as a 
source of information to detect and predict events in real world has been 
limited to the work by Osborne et al. \cite{Osborne2012}, in which
they used Wikipedia page views to fine-filter the outcome of their 
algorithm for Twitter-based ``first story detection'' and a very recent work by Georgescu et al., in which Wikipedia edits 
are introduced as ``entity-specific news tickers and time-lines'' generators \cite{Georgescu2013}. And finally in an interesting work published later than the first
revision of 
the current manuscript, Moat et al. reported on the predictive power of Wikipedia data for financial fluctuations \cite{Moat2013}.

In this work we consider both the activity level of editors and the number of page 
views by readers to assess the popularity of a movie. We define different 
predictor variables and apply a linear regression model
to forecast the first weekend box office revenue of a set of 
312 movies, which were released in the United States in 2010. 
Our analysis not only outperforms the previous works by the much larger number of movies we have investigated, but also improves on the 
state of the art by providing reasonable 
predictions as early as one month prior to the release date of the movie.
Finally, our statistical approach, free of any language 
based analysis, e.g., sentiment analysis, 
can be easily generalized to non-English speaking movie markets 
or even other kinds of products.

\section*{Results}
\begin{figure}[ht!]
  \centering
\includegraphics[width= 0.90 \linewidth]{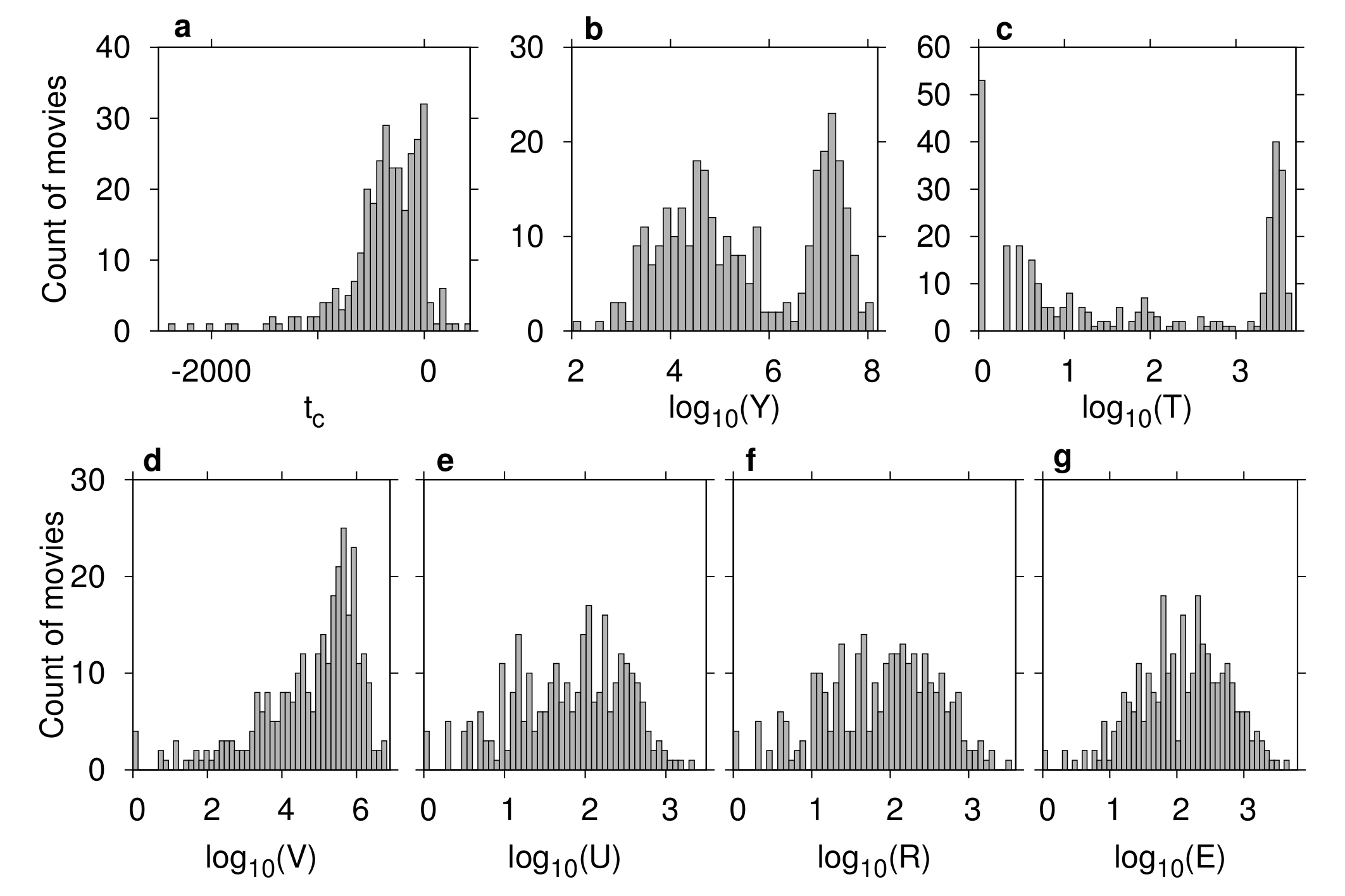}
  \caption{{\bf Histograms of different variables for our sample of $\boldsymbol{n=312}$ movies from 2010.} 
A: Time of creation $t_c$ of the corresponding article in Wikipedia, shown in days of \emph{movie time} ($t=0$ is the release time), 
B: Release weekend box office revenue in the U.~S., in USD 
C: \emph{number of theaters} that screened the movie on the first weekend,
D: Accumulated \emph{number of views}, and
E: \emph{users},
F: \emph{edits},
G: \emph{rigor}
 for the Wikipedia page up to $t=7$ days after release.}
\label{fig:histograms}
\end{figure}

According to data from Box Office Mojo, 
there were 535 movies that were screened in the United States in 2010 (see the Methods section). 
We could track the corresponding page in Wikipedia for 312 of them. 
A closer look at the history of these 312 articles shows that many of them are 
created a lot earlier than the release date of the movie (Fig.~\ref{fig:histograms}(A)). This enables us to
follow the popularity of the movie much in advance. 
To estimate the popularity, we followed four activity measures;
$V$: \emph{Number of views} of the article page,
$U$: \emph{Number of users}, being the number of human editors who have contributed to the article,
$E$: \emph{Number of edits} made by human editors on the article, and $R$: \emph{Collaborative rigor} (or 
simply \emph{rigor} \cite{Kimmons11}) of the editing train of the article. 
To have a consistent time framework, we set the release time of the movie as $t=0$.
For more details see the Methods section. 
Examples of the daily increments of number of views and number of users are shown 
in Supplementary Fig.~S1. The daily increments of both variables rise and fall around the day of release
similarly to observations by Ishii et al. \cite{Ishii12}.
In addition to these, an essential parameter for predicting the 
movie revenue is \emph{the number of theaters} that
screen the movie $T$, which is included in our set of parameters. 
The complete dataset including the financial data as well as Wikipedia activity records
is available via the Supplementary Data S1.
To have an overall image of the sample, histograms of the accumulated 
values of the 4 activity parameters from the first edit on the article up to 7 days 
after release, along with the first weekend box office revenue, and the
number of theaters screening the movie are depicted in Fig.~\ref{fig:histograms}(B--F).
It is clear that revenues among the sample have a bimodal 
distribution (Fig.~\ref{fig:histograms}(B)). This is in accord with \cite{Pan2010}, where authors report that the distribution of the total revenue of a sample
of 5,222
movies released over the period of 1999-2008 across theaters in the USA, 
exhibits bimodal nature and
have been fit using a superposition of two log-normal distributions. It also shows
that Wikipedia coverage is not limited to financially successful movies. 
The considerable amount of activity on Wikipedia articles    
(Fig.~\ref{fig:histograms}(D--G)) indicates the richness of the
data. However, before building a regression model, the correlations 
between the activity parameters and the box office revenue should 
be examined first.

\begin{figure}[ht!]
  \centering
 \includegraphics[width= 0.65 \linewidth]{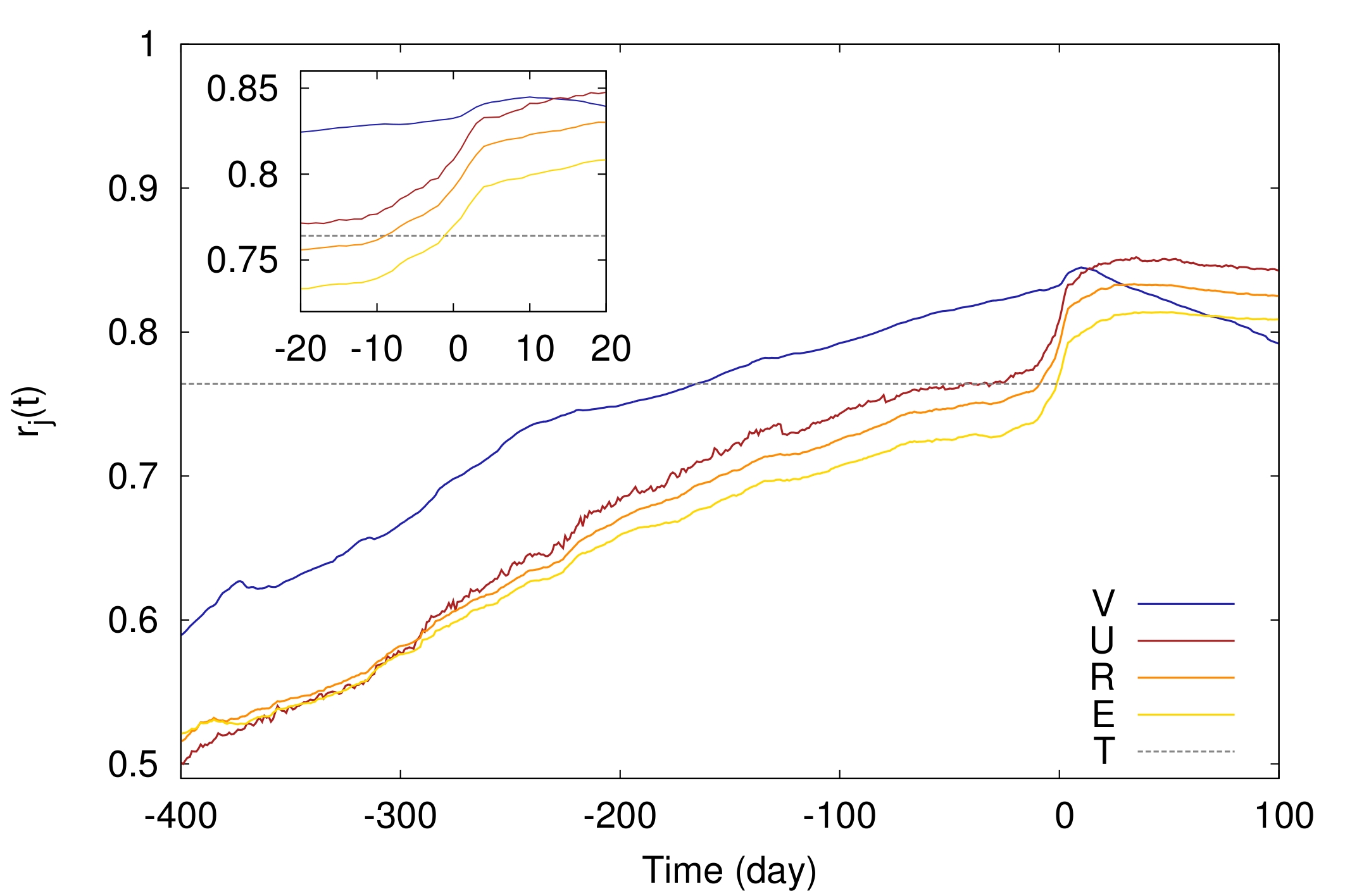}
 \caption{{\bf Temporal evolution of $\boldsymbol{r_j(t)}$, the Pearson correlation of the box office 
revenue with different predictors.} The shorthands $V$, $U$, $R$, $E$, and $T$ denote the \emph{number of views}, 
the \emph{number of users}, the \emph{rigor}, the \emph{number of edits}, and the \emph{number of theaters}, 
respectively. Time is measured in movie time. \emph{Inset:} 
magnified detail of the main panel, showing the Pearson correlation around the day of release. 
Dashed horizontal line shows the correlation for \emph{the number of theaters.}}
  \label{fig:correlations}
\end{figure}

The Pearson correlation coefficient $r_j(t)$ between the accumulated value $x_j(t)$ 
of the $j$-th predictor variable from the inception of the article up to time $t$ before the movie release 
and the box office revenue $y$ is calculated as
\begin{equation}
  \label{eq:pearson}
  r_j(t)=\frac{\langle x_j(t) y \rangle - \langle x_j(t) \rangle \langle y\rangle}{\sqrt{\langle x_j^2(t) \rangle-\langle x_j(t) \rangle^2} 
\sqrt{\langle y^2 \rangle-\langle y\rangle^2}}\text{,}
\end{equation}
with $\langle .\rangle$ indicating average over the whole sample.
Temporal correlations are shown in Fig.~\ref{fig:correlations}.
For all activity based predictors the correlation coefficient gradually increases as time 
approaches the day of release and around the day of release, 
correlation suddenly rises. Note that $V$
shows the highest correlation with the revenue prior to the release pf movies.

\begin{figure}[ht!]
  \centering
  \includegraphics[width= 0.65 \linewidth]{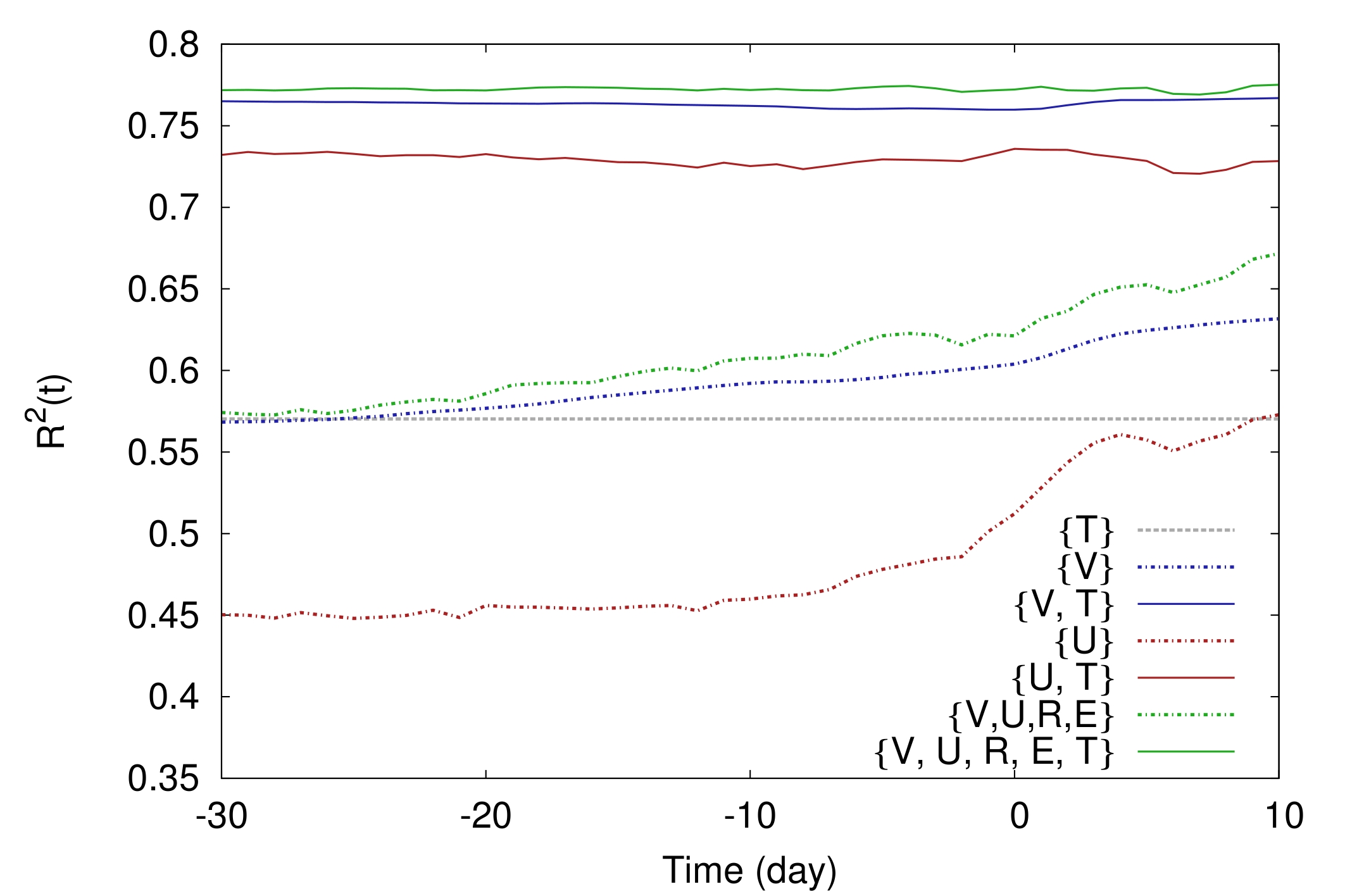}
  \caption{{\bf Coefficient of determination of the multivariate linear regression model 
fed by different set of input variables.}
The shorthands $V$, $U$, $R$, $E$, and $T$ denote the \emph{number of views}, 
the \emph{number of users}, the \emph{rigor}, the \emph{number of edits}, and the \emph{number of theaters}, 
respectively. The coefficient of determination was calculated using 10-fold cross-validation (see the Methods section).
 The dashed gray line shows the coefficient of determination for linear regression solely based on the \emph{number of theaters}.}
  \label{fig:combinations}
\end{figure}

We build a multivariate linear regression model for predicting 
the box office revenue $y$. The general form of a regression model 
at time $t$ before release, based on a set of predictor variables $S$ is
\begin{equation}
  \label{eq:regression}
  y=\sum_{j\in S}\alpha_j(t) x_j(t)+C_S(t)+\varepsilon_S(t)\text{,}     
\end{equation}
where $\alpha_j(t)$s are time varying parameters of the linear regression model, 
$C_S(t)$ is a constant and $\varepsilon_S(t)$ is the noise term. 
We feed the model with different combinations of predictor variables 
and characterize the goodness of different sets by calculating the coefficient of 
determination $R^2(t)$. The coefficient of determination is calculated using 10-fold 
cross-validation (See Methods section). 
Temporal evolution of $R^2(t)$ is shown for different predictor sets $S$ in Fig.~\ref{fig:combinations}. 
While a model employing $\{T\}$ can be seen as a benchmark of the state of the art in real market predictions,
the model solely fed by $\{V\}$ predicts roughly as well as that.
Combinations of $\{V,T\}$ and $\{U,T\}$ score well above the benchmark indicating 
the relevance of activity measures for prediction.
Among all sets considered (not shown here), $\{V,U,R,E,T\}$ yields the highest 
coefficient of determination, which reaches 0.77 around a month before the movie release. 

\section*{Discussion}
Results presented above clearly show how simple use of user 
generated data in a social environment like Wikipedia can enhance our ability to predict the
collective reaction of society to a cultural product. While these results can be of practical application for marketing purpose, 
especially in combination with other source of 
information, our main aim is to demonstrate the extent of engagement of members of the public in the peer-production platforms. 
The introduced approach can be easily generalized to 
other fields where mining of public opinion provides valuable insights, e.g., financial decisions, policy making, and governance. We believe that 
Wikipedia and similar mass-collaboration platforms can serve as alternative resources
for social media streams with higher level of professionalism and deeper engagement of users. 
Since the methods presented here are independent of the language of the medium, 
they can be easily generalized to other languages and local markets.

It is worth mentioning that to feed our predictive model, we have tried several other activity measures, which can potentially be predictive 
parameters, e.g., time span between the creation of the article and the release time and length of the article. However these quantities did not show any significant 
correlation with the box office revenue and consequently were excluded from the model.

\begin{figure}[ht!] 
  \centering
  \includegraphics[width= 0.5 \linewidth]{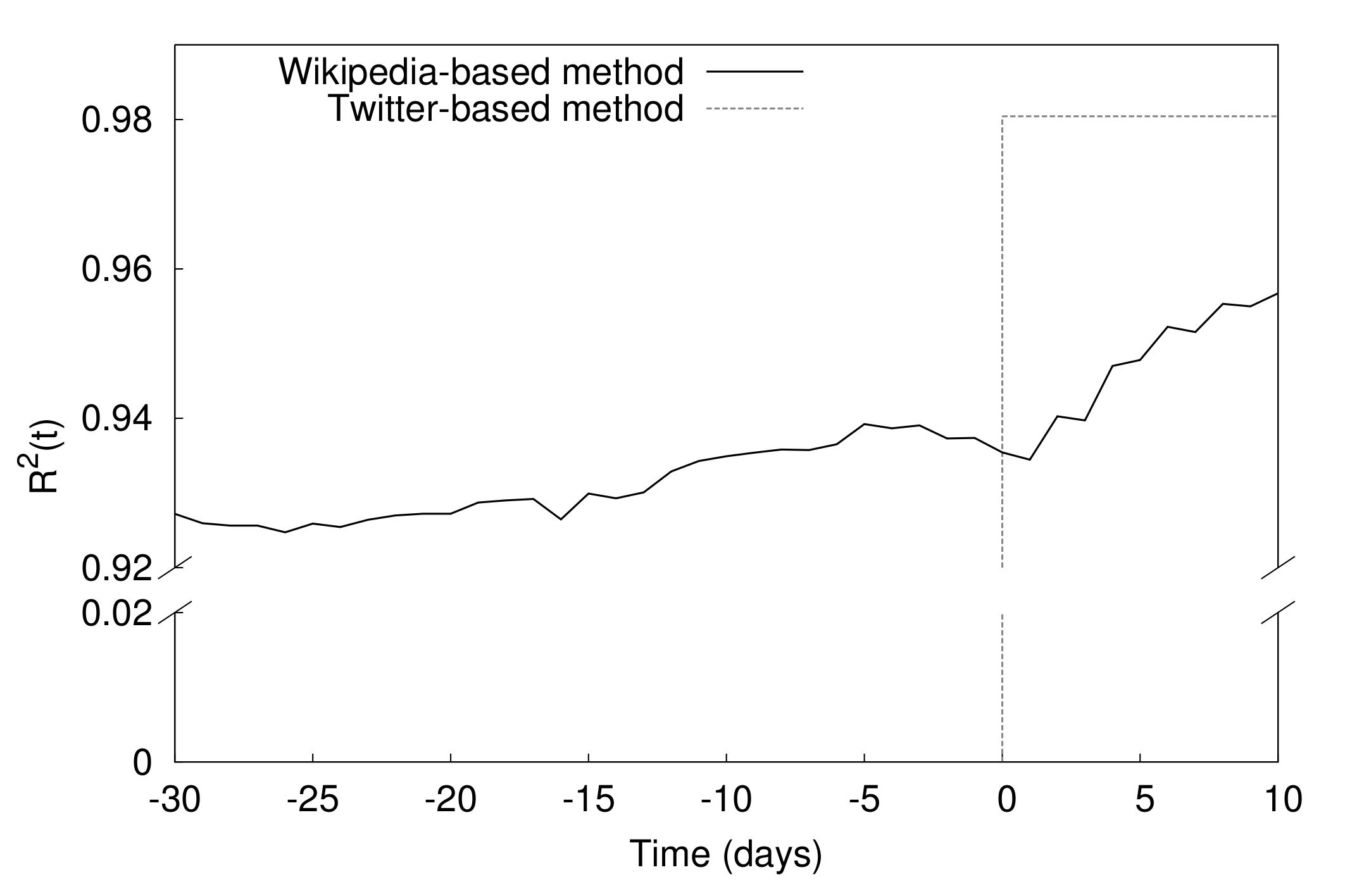}
  \caption{{\bf Comparison of the results with the Twitter-based prediction in Asur and Huberman work \cite{Asur10}.} 
Same sample of 24 movies is considered
 as both training and test set. The coefficient of 
determination obtained with the Twitter-based method is 0.98 at the night 
of the release (day 0 in movie time).}
  \label{fig:comparison}
\end{figure}

We also compare the predictive model based on Wikipedia activity measures with the results of
the Twitter-based model provided in the 2010 study of Asur and Huberman \cite{Asur10}. 
Asur~and~Huberman use a sample of 24 movies to train and test 
their model. In the same approach we train and test our model focusing on the same set of movies.
The $R^2(t)$ of our Wikipedia model reaches $0.94$ few days before release, while 
it is $0.98$ for the Twitter model.
However, the results of the Twitter study are limited to the night before release, while the analysis presented here can 
make predictions with reasonable accuracy ($R^2 > 0.925$) as 
early as one month before release (See Fig.~\ref{fig:comparison}). 
One should also bear in mind that the Wikipedia model does 
not require any complex content analysis and only relies on statistical measures of activity level.
The predicting power of the Wikipedia-based model, despite its simplicity compared to the Twitter, can be explained by the fact that many of the Wikipedia editors
are committed followers of movie industry who gather information and edit related articles significantly earlier than the release date, whereas the ``mass'' production of
tweets only occurs very close to the release time, mostly evoked by marketing campaigns.

\begin{figure}[ht!]
  \centering
  \includegraphics[width= 0.5 \linewidth]{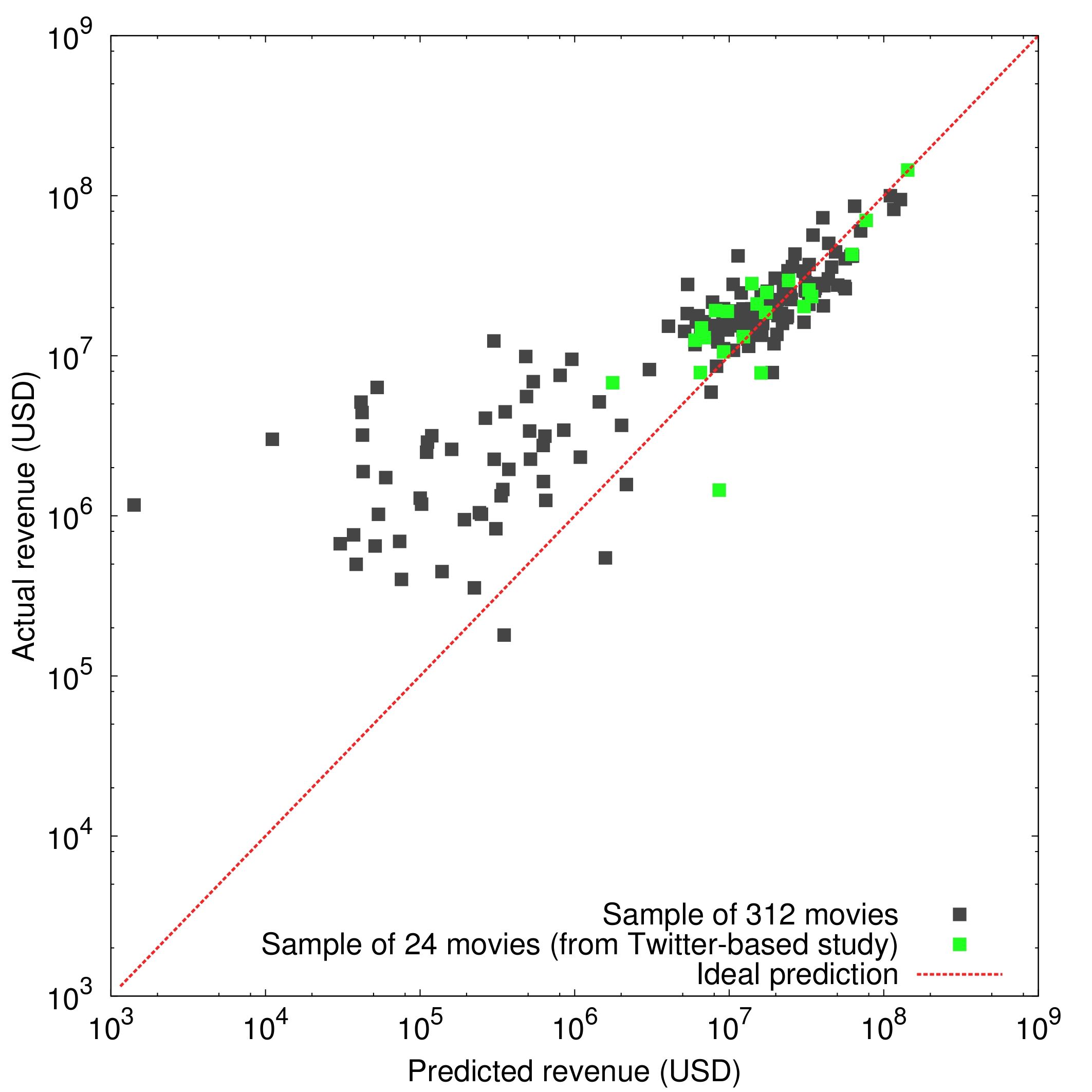}
  \caption{{\bf First weekend box office revenue in the U.~S. against its predicted value by the Wikipedia model at $\boldsymbol{t=-30}$ days.} 
Green dots are representing the smaller sample of 24 movies common in Twitter and Wikipedia 
studies, and black dots are movies from the 2010 sample of 312 movies. 
Note that negative predicted revenues for some of the very unpopular movies could not be shown in the 
logarithmic scale.
}  \label{fig:predicted_vs_real}
\end{figure}

Fig.~\ref{fig:predicted_vs_real} shows the actual revenue of movies 
in the sample against the predicted revenue at $t=-30$ days. 
It is evident that the prediction is more precise for more successful movies. When less successful 
movies are considered, deviations from the diagonal line denoting perfect prediction,
increase. Some examples of the movies whose box office receipts were predicted accurately are
{\it Iron Man 2}, {\it Alice in Wonderland}, {\it Toy Story 3}, {\it Inception}, {\it Clash of the Titans}, and  {\it Shutter Island}. However,
the model failed to
provide accurate predictions for less successful movies, e.g., {\it Never Let Me Go}, {\it Animal Kingdom}, {\it The Girl on the Train}, {\it The Killer Inside Me}, and {\it The Lottery}.
This systematic difference in precision can be explained by the amount of data available for each class of movies. Clearly the model works more accurately when the movie is
more popular and the volume of the related data is larger. 
By considering the green squares which represent the movies in the sample predicted by the Twitter model, one realizes that 
most of the movies predicted by the Twitter method are among  
the successful ones, therefore applicability of the Twitter model on movies with medium and low popularity levels 
remains an open question. 

While we tried to keep our model as simple as possible and based on only a 
few variables, one could possibly enhance the efficiency of prediction by 
applying more sophisticated statistical methods, such as neural 
networks on more detailed content-related parameters e.g., the controversy
measure of the article \cite{sumi2011b}.

\FloatBarrier
\section*{Methods}
In this study we consider a sample of 
312 movies, which were released in the United States in 2010.
The complete dataset including the financial data as well as Wikipedia activity records
is available via the Supplementary Data S1.
To obtain this dataset, first the list of 2010 movies distributed 
in the U.~S. is acquired from Box Office Mojo (\url{http://boxofficemojo.com}) 
along with their accompanying financial data (535 movies). 
Financial data consist of the opening weekend box office 
revenue and the number of theaters screening the movie.

In order to locate the corresponding articles in Wikipedia, 
we use the category system of Wikipedia. Wikipedia articles
are classified into one or more categories by users.
We match the title of 
the movies in the Mojo database with the title of Wikipedia pages in 
categories {\tt 2009 films} and {\tt 2010 films}. 
Inclusion of the category {\tt 2009 films} is necessary because of movies that were released 
in 2010 in the U.S. but which could have already entered the international market during 2009, and hence were 
classified in the category {\tt 2009 films} in Wikipedia. 
To achieve the best possible match of the 
titles, they were stripped of punctuation and postfixes. Wikipedia uses the latter to maintain the uniqueness 
of every title, such as in the case of {\tt Avatar (2009 film)} and {\tt Avatar (computing)}. 
As a result of the matching process described above, a sample consisting of the financial data and the 
corresponding Wikipedia page for 312 movies was obtained. 

For the sake of convenience we introduce 
\emph{movie time}, a common time coordinate for the 
movies in the scope of our study. By definition, movie 
time is measured from the time of release 
in the U.S. All temporal variables 
are measured in movie time. Throughout this study, we
consider accumulated values of parameters from the inception of the 
article to the prediction time $t$ for each activity measure.
The four activity measures are defined as the following:

\emph{Number of users, $U$}: the number of different human 
users who contributed to the page. 

\emph{Number of edits, $E$}: the number of modifications made 
by human users on the article. 

\emph{Collaborative rigor, $R$}: similar to the number of edits; however it counts multiple 
subsequent edits by the same user as one edit \cite{Kimmons11}. 
It avoids counting multiple edits by the same user 
in a short period, e.g., to correct errors in their previous contribution. 

\begin{figure}[ht!]
   \centering
   \includegraphics[width= 0.4 \linewidth]{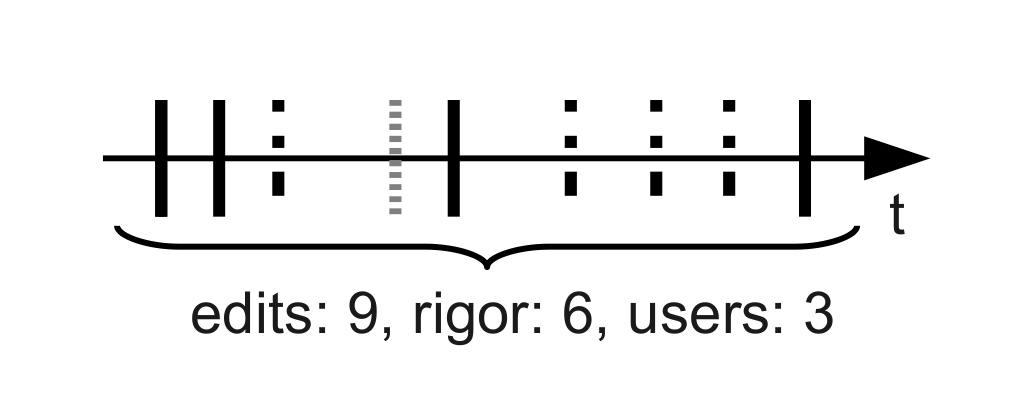} 
   \caption{{\bf Illustration of different variables characterizing the activity of 
Wikipedia editors on an article.} Each tick on the axis represents 
a modification of the page. Different tick styles refer to different users.}
   \label{fig:observables}
\end{figure}

A schematic illustration of these activity measures is presented in  Fig.~\ref{fig:observables}.  
These three variables are calculated using the page history databases of Wikimedia 
Toolserver (\protect\url!http://toolserver.wikimedia.org!), which register information 
about every modification made to the pages of Wikipedia. To ensure that the above variables count 
solely human activity, contributions made by \emph{bots} are excluded from calculations. 
Bots are automated scripts which facilitate automatic tasks such as spell checking. Contributions 
made by bots are registered in the same way as revisions by humans; however, they can be 
distinguished from human activity by noting a special entry in the databases of Wikimedia Toolserver, 
called the \emph{bot flag}.

\emph{Number of views, $V$}: the number of times a given page is viewed 
from its inception up to the time $t$. 
This data is extracted from the page view statistics section of the Wikimedia Downloads 
site (\protect\url!http://dumps.wikimedia.org/other/pagecounts-raw!) through 
the web-based interface of ``Wikipedia article traffic statistics'' (\protect\url!http://stats.grok.se!).
Wikimedia Downloads counts views 
only since December 2007 and the view count data for July 2008 
is corrupted. Therefore it is 
impossible to count the exact total number of views till the time of 
prediction for all considered pages. 
We have counted the page hits from $t=-500$ days before release, 
which according to Fig.~\ref{fig:histograms}(A), is sufficiently early.
Another challenge is created by the renaming of the articles, which splits page hit counts into subsets 
according to the various titles the page possesses throughout its history. To cope with this problem, we followed the logs of ``title moves'' in the article history
to track back and merge the whole page hits. Note that in the the dataset there are records on Wikipedia page requests for non-existing pages as well, which give us 
an indicator of the public interest in a movie even before its Wikipedia article is created and therefore we did not exclude such records from the data. 
\emph{Number of theaters}: the count of movie theaters that screen the movie on the first 
weekend of its release.

To calculate the coefficient of determination, 
we carry out 10-fold cross-validation by randomly dividing our 
sample of 2010 movies into 10 subsets first. In the next step the model is 
trained for the union of the 9 subsets and tested on the remaining 10th subset.
This is repeated for all 10 permutations of the subsets and the coefficient of determination for the
model is obtained as the average over the permutations.

\FloatBarrier
\section*{Acknowledgments}
We thank Wikimedia Deutschland e.V. for providing access to
its databases on the Wikimedia Toolserver and IMDb, Inc. for
the access to Box Office Mojo database. We also thank the PLoS ONE anonymous reviewers for useful comments.
Partial financial support from EU's 7th Framework Program's FET-Open to ICTeCollective project no. 238597 and by the Academy of Finland, 
the Finnish Center of Excellence program, project no. 129670, and TEKES (FiDiPro) are gratefully acknowledged.

\bibliography{wikipredict}

\FloatBarrier
\section*{Supporting Information}
\setcounter{figure}{0}            
\makeatletter
\renewcommand{\thefigure}{S\@arabic\c@figure}
\begin{figure}[ht!]
  \centering
\includegraphics[width= 0.85 \linewidth]{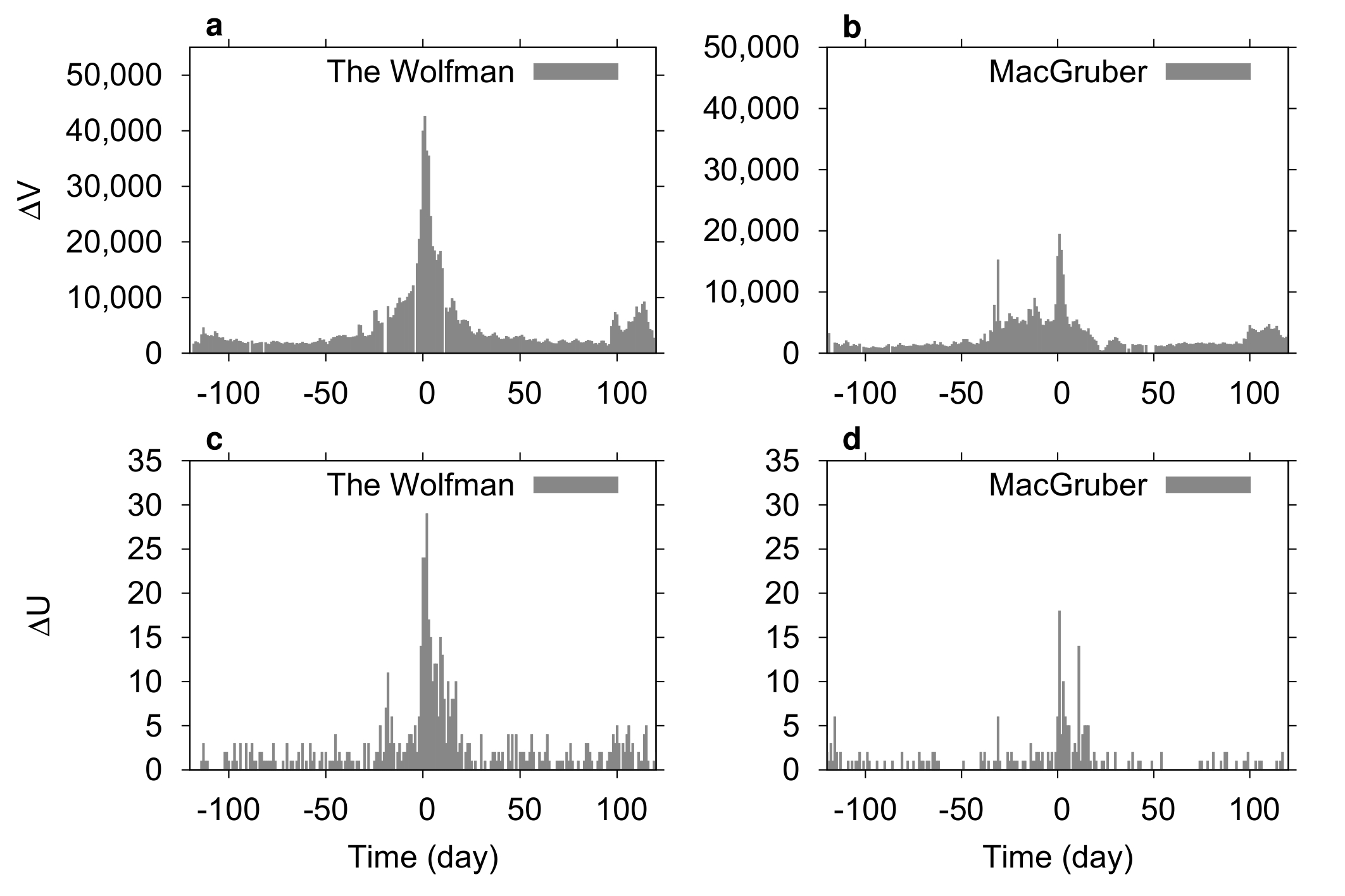}
 \caption{{\bf Temporal evolution of Wikipedia-based predictors for two individual movies: 
{\tt  The Wolfman (2010)} and {\tt MacGruber}.} The daily increments of \emph{number of views} $\Delta V$
and \emph{number of users} $\Delta U$ are shown for the articles in English Wikipedia that correspond to the two movies. 
The temporal axis shows movie time, i.e., a time-frame in which $t=0$ corresponds to the release date. {\tt The Wolfman} earned a box 
office revenue of \$$31,479,235$ on the release weekend while {\tt MacGruber} gained only \$$4,043,495$. 
Accordingly, predictor variables take larger values in the case of {\tt The Wolfman}.}
\label{fig:individual}
\end{figure}

\subsubsection*{Dataset S1}
The dataset under study, including the financial and Wikipedia activity data is also available at\\ \url{http://wwm.phy.bme.hu/SupplementaryDataS1.zip}

\end{document}